\documentclass[aps,prl,twocolumn,superscriptaddress,showpacs,amsmath,amssymb,floatfix]{revtex4}

\usepackage{graphicx}
\usepackage{dcolumn}
\usepackage{bm}
\usepackage{url} 
\usepackage{amsmath} 
\usepackage{amssymb} 

\begin{document}

\title{\quad\\[1.0cm] 
Search for $CP$ Violation in the Decays $D^0\rightarrow K^0_S P^0$
}

\affiliation{Budker Institute of Nuclear Physics, Novosibirsk}
\affiliation{Faculty of Mathematics and Physics, Charles University, Prague}
\affiliation{University of Cincinnati, Cincinnati, Ohio 45221}
\affiliation{Gifu University, Gifu}
\affiliation{Gyeongsang National University, Chinju}
\affiliation{Hanyang University, Seoul}
\affiliation{University of Hawaii, Honolulu, Hawaii 96822}
\affiliation{High Energy Accelerator Research Organization (KEK), Tsukuba}
\affiliation{Indian Institute of Technology Guwahati, Guwahati}
\affiliation{Institute of High Energy Physics, Vienna}
\affiliation{Institute of High Energy Physics, Protvino}
\affiliation{Institute for Theoretical and Experimental Physics, Moscow}
\affiliation{J. Stefan Institute, Ljubljana}
\affiliation{Kanagawa University, Yokohama}
\affiliation{Institut f\"ur Experimentelle Kernphysik, Karlsruher Institut f\"ur Technologie, Karlsruhe}
\affiliation{Korea Institute of Science and Technology Information, Daejeon}
\affiliation{Korea University, Seoul}
\affiliation{Kyungpook National University, Taegu}
\affiliation{\'Ecole Polytechnique F\'ed\'erale de Lausanne (EPFL), Lausanne}
\affiliation{Faculty of Mathematics and Physics, University of Ljubljana, Ljubljana}
\affiliation{University of Maribor, Maribor}
\affiliation{Max-Planck-Institut f\"ur Physik, M\"unchen}
\affiliation{University of Melbourne, School of Physics, Victoria 3010}
\affiliation{Nagoya University, Nagoya}
\affiliation{Nara Women's University, Nara}
\affiliation{National Central University, Chung-li}
\affiliation{National United University, Miao Li}
\affiliation{Department of Physics, National Taiwan University, Taipei}
\affiliation{H. Niewodniczanski Institute of Nuclear Physics, Krakow}
\affiliation{Nippon Dental University, Niigata}
\affiliation{Niigata University, Niigata}
\affiliation{University of Nova Gorica, Nova Gorica}
\affiliation{Novosibirsk State University, Novosibirsk}
\affiliation{Osaka City University, Osaka}
\affiliation{Panjab University, Chandigarh}
\affiliation{Saga University, Saga}
\affiliation{University of Science and Technology of China, Hefei}
\affiliation{Seoul National University, Seoul}
\affiliation{Sungkyunkwan University, Suwon}
\affiliation{School of Physics, University of Sydney, NSW 2006}
\affiliation{Tata Institute of Fundamental Research, Mumbai}
\affiliation{Excellence Cluster Universe, Technische Universit\"at M\"unchen, Garching}
\affiliation{Toho University, Funabashi}
\affiliation{Tohoku Gakuin University, Tagajo}
\affiliation{Tohoku University, Sendai}
\affiliation{Tokyo Institute of Technology, Tokyo}
\affiliation{Tokyo Metropolitan University, Tokyo}
\affiliation{Tokyo University of Agriculture and Technology, Tokyo}
\affiliation{CNP, Virginia Polytechnic Institute and State University, Blacksburg, Virginia 24061}
\affiliation{Yonsei University, Seoul}
  \author{B.~R.~Ko}\affiliation{Korea University, Seoul} % Korea
  \author{E.~Won}\affiliation{Korea University, Seoul} % Korea
  \author{K.~Arinstein}\affiliation{Budker Institute of Nuclear Physics, Novosibirsk}\affiliation{Novosibirsk State University, Novosibirsk} % BINP
  \author{T.~Aushev}\affiliation{\'Ecole Polytechnique F\'ed\'erale de Lausanne (EPFL), Lausanne}\affiliation{Institute for Theoretical and Experimental Physics, Moscow} % ITEP
  \author{A.~M.~Bakich}\affiliation{School of Physics, University of Sydney, NSW 2006} % Sydney
  \author{V.~Balagura}\affiliation{Institute for Theoretical and Experimental Physics, Moscow} % ITEP
  \author{E.~Barberio}\affiliation{University of Melbourne, School of Physics, Victoria 3010} % Melbourne
  \author{K.~Belous}\affiliation{Institute of High Energy Physics, Protvino} % Protvino
  \author{V.~Bhardwaj}\affiliation{Panjab University, Chandigarh} % Panjab
  \author{B.~Bhuyan}\affiliation{Indian Institute of Technology Guwahati, Guwahati} % IITG
% ...
% BINP
\collaboration{The Belle Collaboration}
 
\begin{abstract}
We have searched for $CP$ violation in the decays $D^0\rightarrow K^0_S P^0$
where $P^0$ denotes a neutral pseudo-scalar meson that is either a $\pi^0$,
$\eta$, or $\eta'$ using KEKB asymmetric-energy $e^+e^-$ collision data
corresponding to an integrated luminosity of 791 fb$^{-1}$ collected with the
Belle detector. No evidence of significant $CP$ violation is observed. We
report the most precise $CP$ asymmetry measurement in the decay $D^0\rightarrow
K^0_S\pi^0$ to date: $A_{CP}^{D^0\rightarrow
  K^0_S\pi^0}=(-0.28\pm0.19\pm0.10)\%$. We also report the first measurements
of $CP$ asymmetries in the decays $D^0\rightarrow K^0_S\eta$ and
$D^0\rightarrow K^0_S\eta'$: $A_{CP}^{D^0\rightarrow
  K^0_S\eta}=(+0.54\pm0.51\pm0.16)\%$ and $A_{CP}^{D^0\rightarrow
  K^0_S\eta'}=(+0.98\pm0.67\pm0.14)\%$, respectively.

\end{abstract}
\maketitle

{\renewcommand{\thefootnote}{\fnsymbol{footnote}}}
\setcounter{footnote}{0}

The recent evidence for $D^0-\bar{D}^0$
mixing~\cite{D0MIXING_Babar,D0MIXING_Belle,D0MIXING_CDF} and the corresponding
mixing parameters~\cite{HFAG} are at the upper edge of standard model (SM)
predictions~\cite{DMIX}. However, large theoretical uncertainties in these
predictions limit the sensitivity to effects of physics beyond the SM. An
alternative, potentially more promising approach to search for new physics (NP)
is the study of violation of the combined Charge-conjugation and Parity
symmetries ($CP$) in the decays of charmed mesons~\cite{YAY}. In contrast to
mixing, the expected SM $CP$ violation in the charm sector is
small~\cite{SMCP}.

In this Letter we report time-integrated $CP$ asymmetry measurements in the
decays $D^0\rightarrow K^0_S P^0$~\cite{CC} where $P^0$ denotes a neutral
pseudo-scalar meson: $\pi^0$, $\eta$, or $\eta'$. The time-integrated
asymmetry, $A_{CP}$, is defined as
\begin{equation}
  A^{D^0\rightarrow K^0_S P^0}_{CP}~=~\frac
  {\Gamma(D^0\rightarrow K^0_S P^0)-\Gamma(\bar{D}^0\rightarrow K^0_S P^0)}
  {\Gamma(D^0\rightarrow K^0_S P^0)+\Gamma(\bar{D}^0\rightarrow K^0_S P^0)},
\label{EQ:ACP}
\end{equation}
where $\Gamma$ is the partial decay width. 

The observed $K^0_S P^0$ final states are mixtures of $D^0\rightarrow\bar{K}^0
P^0$ and $D^0\rightarrow K^0 P^0$ decays where the former are Cabibbo-favored
(CF) and the latter are doubly Cabibbo-suppressed (DCS). In the absence of
direct $CP$ violation in CF and DCS decays, as expected in the SM, the $CP$
violation in these processes within the SM is generated from mixing and
interference of decays with and without mixing, which is parameterized by
$a^{\rm ind}$ (we adopt the symbols used in Ref.~\cite{YAY}). SM
$K^0-\bar{K}^0$ mixing leads to a small $CP$ asymmetry in final states
containing a neutral kaon, even if no $CP$ violating phase exists in the charm
decay. The asymmetry that is expected from the SM is measured to be
$(-0.332\pm0.006)$\%~\cite{PDG2010} from $K^0_L$ semileptonic decays and
referred to as $A^{\bar{K}^0}_{CP}$~\cite{ACPK0S}, which is reflected in the
value of $A^{D^0\rightarrow K^0_S P^0}_{CP}$ if DCS decay contributions are
ignored. Since the $a^{\rm ind}$ value expected from the SM is at most
$\mathcal{O}(10^{-4})$~\cite{YAY,SMCP}, the value of $CP$ asymmetry in the
decays $D^0\rightarrow K^0_S P^0$ within the SM is approximately
$A^{\bar{K}^0}_{CP}$. On the other hand, if NP processes contain additional
weak phases other than the one in the Kobayashi-Maskawa ansatz~\cite{KM},
interferences between CF and DCS decays could generate $\mathcal{O}(1)$\%
direct $CP$ asymmetry in the decays $D^0\rightarrow K^0_S
P^0$~\cite{BIGI}. Physics beyond the SM could also induce $\mathcal{O}(1)$\%
indirect $CP$ asymmetry~\cite{YAY}. Thus, observing $A_{CP}$ inconsistent with
$A^{\bar{K}^0}_{CP}$ in $D^0\rightarrow K^0_S P^0$ decays would be strong
evidence for processes involving physics beyond the SM~\cite{YAY,BIGI}.

In addition to $A_{CP}$ measurements, we examine the universality of $a^{\rm
  ind}$ in $D^0$ decays~\cite{YAY} by comparing our previous
result~\cite{D0MIXING_Belle} with the $A^{D^0\rightarrow K^0_S\pi^0}_{CP}$
value reported in this Letter. Our previously measured values of direct $CP$
violation asymmetries (denoted $a^d_f$~\cite{YAY}), $a^d_{D^0\rightarrow
  K^+K^-}$ and $a^d_{D^0\rightarrow\pi^+\pi^-}$~\cite{D0hhBelle} are also
updated.

The decay $D^{*+}\rightarrow D^0\pi^+_s$ is used to identify the flavor of the
$D^0$ meson from the charge of the low momentum pion (referred to as ``the soft
pion''), $\pi^+_s$. Thus, we determine $A^{D^0\rightarrow K^0_S P^0}_{CP}$ by
measuring the asymmetry in the signal yield
\begin{equation}
  A^{D^{*+}\rightarrow D^0\pi^+_s}_{\rm rec}=\frac
  {N_{\rm rec}^{D^{*+}\rightarrow D^0\pi^+_s}-N_{\rm rec}^{D^{*-}\rightarrow\bar{D}^0\pi^-_s}}
  {N_{\rm rec}^{D^{*+}\rightarrow D^0\pi^+_s}+N_{\rm
      rec}^{D^{*-}\rightarrow\bar{D}^0\pi^-_s}},    
  \label{EQ:ARECONI}
\end{equation}
where $N_{\rm rec}$ is the number of reconstructed decays. The measured
asymmetry in Eq.~(\ref{EQ:ARECONI}) includes two contributions other than
$A_{CP}$. One is the forward-backward asymmetry ($A_{FB}$) due to
$\gamma^{*}-Z^0$ interference in $e^+e^-\rightarrow c\bar{c}$ and the other is
a detection efficiency asymmetry between positively and negatively charged soft
pions ($A^{\pi^+_s}_{\epsilon}$). Since we reconstruct the $K^0_S$ with
$\pi^+\pi^-$ combinations and $P^0$ with the $\gamma\gamma$ or
$\gamma\gamma\pi^+\pi^-$ final states, asymmetries in $K^0_S$ and $P^0$
detection cancel out. Equation~(\ref{EQ:ARECONI}) then can be simplified to
give
\begin{equation} 
  \begin{split}
    &A^{D^{*+}\rightarrow D^0\pi^+_s}_{\rm rec}~=\\
    &A^{D^0\rightarrow K^0_S P^0}_{CP}+A^{D^{*+}}_{FB}(\cos\theta^{\rm CMS}_{D^{*+}})+A^{\pi^+_s}_{\epsilon}(p^{\rm lab}_{T\pi^+_{s}},\cos\theta^{\rm lab}_{\pi^+_s})
  \end{split}
  \label{EQ:ARECONII}
\end{equation}
by neglecting the terms involving the product of asymmetries, where $A_{CP}$ is
independent of all kinematic variables, $A^{D^{*+}}_{FB}$ is an odd function of
the cosine of the polar angle of $D^{*+}$ in the center-of-mass system (CMS),
and $A^{\pi^+_s}_{\epsilon}$ depends on transverse momentum and polar angle of
$\pi^+_s$ in the laboratory frame, while it is uniform in azimuthal angle. To
correct for $A^{\pi^+_s}_{\epsilon}$ we use the decays $D^0\rightarrow
K^-\pi^+$ (referred to as untagged) and $D^{*+}\rightarrow
D^0\pi^+_s\rightarrow K^-\pi^+\pi^+_s$ (referred to as tagged), and assumes the
same $A_{FB}$ for $D^{*+}$ and $D^0$ mesons. By subtracting the measured
asymmetries in these two decay modes, $A^{\rm untagged}_{\rm rec}$ and $A^{\rm
  tagged}_{\rm rec}$, we directly measure the $A^{\pi^+_s}_{\epsilon}$
correction factor~\cite{D0hhBelle,D0hhBaBar}. With $A^{D^{*+}\rightarrow
  D^0\pi^+_s}_{\rm rec}$ corrected for $A^{\pi^+_s}_{\epsilon}$
(denoted $A^{D^{*+}\rightarrow D^0\pi^+_{s}}_{\rm rec, corr}$ below),
\begin{equation} 
  A^{D^{*+}\rightarrow D^0\pi^+_{s}}_{\rm rec, corr}~=~A^{D^0\rightarrow K^0_S P^0}_{CP}~+~A^{D^{*+}}_{FB}(\cos\theta^{\rm CMS}_{D^{*+}}),
  \label{EQ:ACPFB0}
\end{equation}
we extract $A_{CP}$ and $A_{FB}$ using
\begin{subequations}
  \begin{equation} 
    \begin{split}
      & A^{D^0\rightarrow K^0_S P^0}_{CP}~=~[A^{D^{*+}\rightarrow D^0\pi^+_{s}}_{\rm rec, corr}(\cos\theta^{\rm CMS}_{D^{*+}}) \\
	&~~~~~~\,~~~~~~~~~+~A^{D^{*+}\rightarrow D^0\pi^+_{s}}_{\rm rec, corr}(-\cos\theta^{\rm CMS}_{D^{*+}})]/2, \\%
    \end{split}
    \label{EQ:ACPFB1}
  \end{equation}
  \begin{equation}
    \begin{split} 
      & A^{D^{*+}}_{FB}~=~[A^{D^{*+}\rightarrow D^0\pi^+_{s}}_{\rm rec, corr}(\cos\theta^{\rm CMS}_{D^{*+}}) \\
	&~~~~~\,~~~-~A^{D^{*+}\rightarrow D^0\pi^+_{s}}_{\rm rec, corr}(-\cos\theta^{\rm CMS}_{D^{*+}})]/2. \\%
    \end{split}
    \label{EQ:ACPFB2}
  \end{equation}
\end{subequations}

The data used in this analysis were recorded at or near the $\Upsilon(4S)$
resonance with the Belle detector~\cite{BELLE} at the $e^+e^-$
asymmetric-energy collider KEKB~\cite{KEKB}. The sample corresponds to an
integrated luminosity of 791 fb$^{-1}$.

We apply the same charged track selection criteria that were used in
Ref.~\cite{BRKS}. For soft pions we do not require associated hits in the
silicon vertex detector, either in the $z$ or radial
directions~\cite{SVD2}. Charged kaons and pions are identified by requiring the
ratio of particle identification likelihoods~\cite{BRKS} to be greater or less
than 0.6, respectively.
$K^0_S$ candidates are reconstructed from pairs of oppositely charged tracks
that have an invariant mass within $\pm$9 MeV/$c^2$ of the nominal $K^0_S$
mass~\cite{PDG2010,BRKS}.
Candidate $\pi^0$ and $\eta$ mesons are reconstructed from $\gamma\gamma$ pairs
where the minimum energy of each $\gamma$ is required to be 60 MeV for the
barrel and 100 MeV for the forward region of the calorimeter~\cite{ECL}. We
require the $\gamma\gamma$ invariant mass to be between 0.11 and 0.16 GeV/$c^2$
for $\pi^0$ candidates and between 0.50 and 0.58 GeV/$c^2$ for $\eta$
candidates. The momentum of $\gamma\gamma$ pairs is required to be greater than
0.5 GeV/$c$ for both $\pi^0$ and $\eta$ selections.
In order to remove a significant $\pi^0$ photon background contribution under
the $\eta$ signal peak, we combine individual $\gamma$ candidates from
$\eta\rightarrow\gamma\gamma$ with any other detected $\gamma$ in the event. If
the $\gamma\gamma$ pair invariant mass is in the $\pi^0$ mass window, the
$\gamma$ is rejected. Further reduction of the $\pi^0$ contribution under the
$\eta$ signal is achieved by requiring the energy balance of the $\gamma\gamma$
in the $\eta$ decay to be less than 0.8, where the energy balance is the ratio
of the difference and the sum of two $\gamma$ energies.
Candidate $\eta'$ mesons are reconstructed in the $\eta\pi^+\pi^-$ decay
channel. To improve the $\eta'$ mass resolution, the four-momentum of the
$\eta$ is recalculated with a nominal $\eta$ mass~\cite{PDG2010}
constraint. The same minimum $\gamma$ energy requirement used for the
$D^0\rightarrow K^0_S\eta$ selection is imposed in the $\eta'$
reconstruction. The $\pi^0$ veto, however, is not applied since it is found to
be unnecessary once the invariant mass of the $\eta\pi^+\pi^-$ candidates is
required to be between 0.945 and 0.970 GeV/$c^2$ and the $D^0$ mass selection
requirement, which is described below, is applied.

\begin{figure}[htbp]
\mbox{
  \includegraphics[height=0.16\textwidth,width=0.52\textwidth]{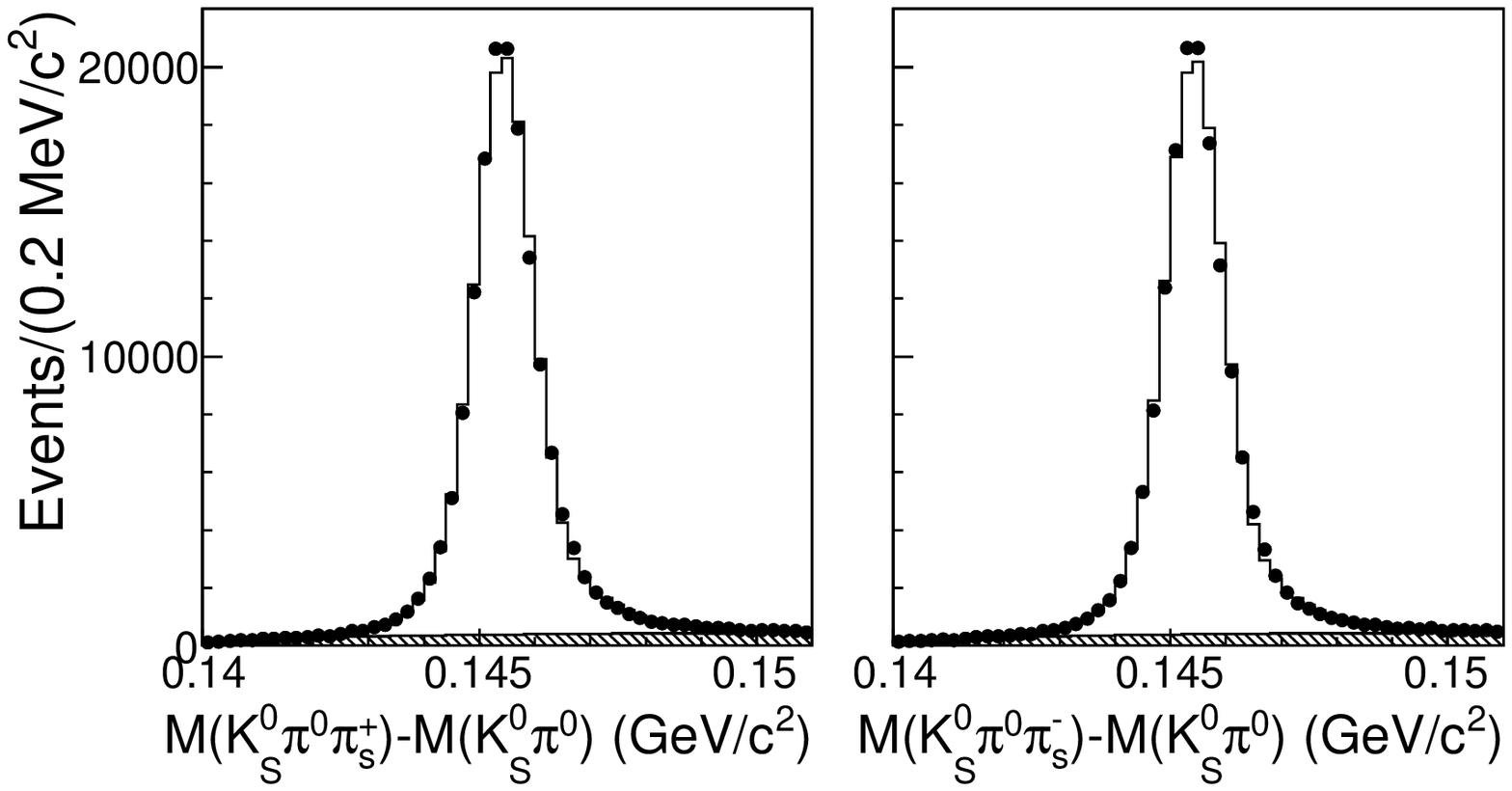}
}
\mbox{
  \includegraphics[height=0.16\textwidth,width=0.52\textwidth]{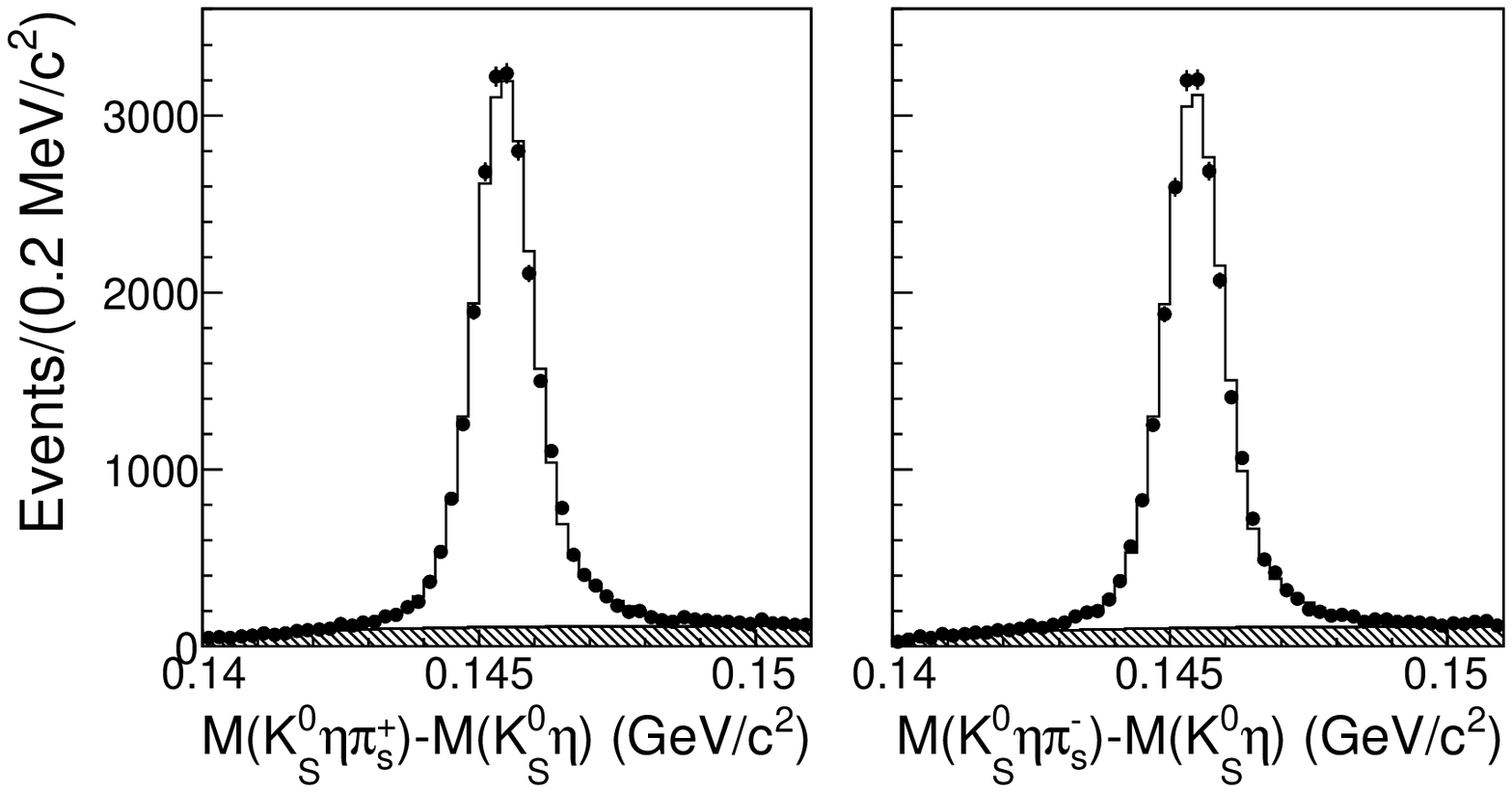}
}
\mbox{
  \includegraphics[height=0.16\textwidth,width=0.52\textwidth]{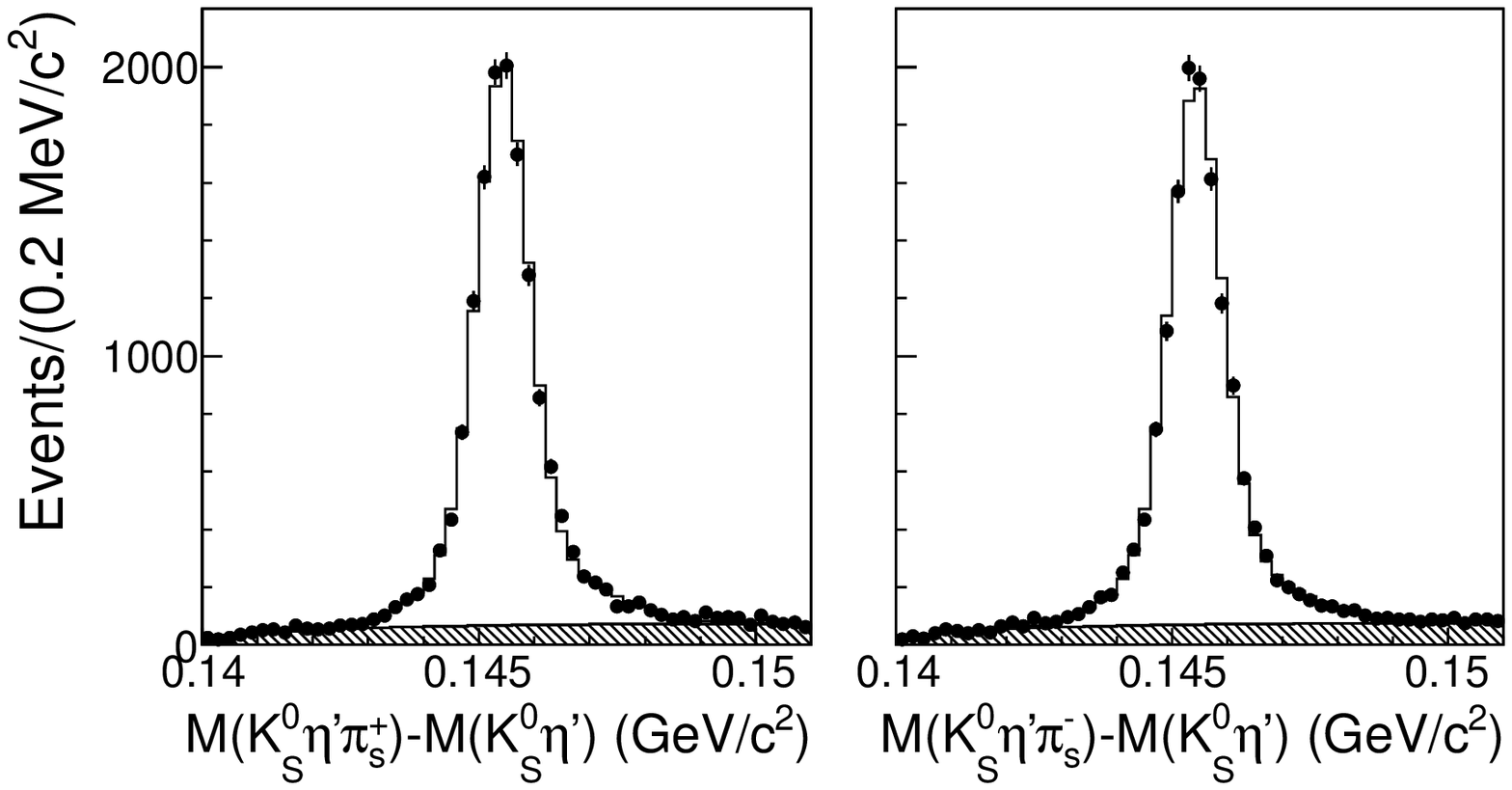}
}
\caption{Distributions of the mass difference $M(D^*)-M(D)$ for the studied
  decay modes. Left plots show the mass difference between $D^{*+}$ and $D^0$
  and right plots show that between $D^{*-}$ and $\bar{D}^0$. Top plots are for
  the $K^0_S\pi^0$, middle plots for the $K^0_S\eta$, and bottom plots for the
  $K^0_S\eta'$ final states. Points with error bars are the data and the
  histograms show the results of the parameterizations of the data. Hatched
  areas are the background contributions.}
\label{FIG:MKSP0}
\end{figure}

The four-momentum of the $P^0$ is recalculated from a kinematic fit to its
nominal mass~\cite{PDG2010} and combined with a $K^0_S$ to form a $D^0$
candidate. $D^{*+}$ candidates are reconstructed using a soft pion and a $D^0$
candidate with mass in the $[1.75, 1.95]$ GeV/$c^2$ $(K^0_S\pi^0)$, $[1.82,
  1.90]$ GeV/$c^2$ $(K^0_S\eta)$, or $[1.84, 1.89]$ GeV/$c^2$ $(K^0_S\eta')$
interval which depends on the mass resolution. To remove $D^{*+}$ mesons produced
in $B$ decays, the $D^{*+}$ momentum in the CMS is required to be greater than
2.5 GeV/$c$. All selections are chosen to maximize $N_S/\sigma_{N_S}$ and to
minimize the peaking backgrounds, where $N_S$ is the signal yield from the fit
and $\sigma_{N_S}$ is the uncertainty in $N_S$. After applying all of the
selections described above, the $D^0\rightarrow\pi^+\pi^-\pi^0$ contribution to
$D^0\rightarrow K^0_S\pi^0$ and the $D^0\rightarrow K^0_S\pi^0$ contribution to
$D^0\rightarrow K^0_S\eta$ are found to be negligible in simulation
studies. Figure~\ref{FIG:MKSP0} shows data distributions of the mass difference
$M(D^{*})-M(D)$ for all the decay modes.

All mass difference signals are parameterized as a sum of a Gaussian and a
bifurcated Gaussian distributions with a common mean. The background is
parameterized by the form $(x-m_{\pi^+})^\alpha e^{-\beta(x-m_{\pi^+})}$, where
$\alpha$ and $\beta$ are free parameters, $m_{\pi^+}$ is the charged pion
mass~\cite{PDG2010} and $x$ is the mass difference. The asymmetry and the sum
of the $D^{*+}$ and $D^{*-}$ yields are directly obtained from a simultaneous
fit to the $D^{*+}$ and $D^{*-}$ candidate distributions. The common parameters
in the simultaneous fit are the mean of the Gaussian, the widths of the
Gaussian and the bifurcated Gaussian, and the ratio of the Gaussian and the
bifurcated Gaussian amplitudes, which are the same for the $M(D^*)-M(D)$
distributions in different $K^0_S P^0$ final states and in the slightly
different phase spaces of the individual $K^0_S P^0$
modes. Table~\ref{TABLE:FIT} lists the results of the fits. 
\begin{table}[htbp]
  \caption{The sum ($N_{S}$) and the asymmetry ($A_{\rm rec}$ in
  Eq.~(\ref{EQ:ARECONI})) of $D^{*+}$ and $D^{*-}$ yields from the fits. The
  uncertainties are statistical only.}
  \label{TABLE:FIT}
  \begin{ruledtabular}
    \begin{tabular}{lcc} 
                                                           &$N_{S}$         &$A_{\rm rec}$ (\%) \\ \hline
$D^{*+}\rightarrow D^0\pi^+_s\rightarrow K^0_S\pi^0\pi^+_s$&$326303\pm679$  &$+0.19\pm0.19$\\ 
$D^{*+}\rightarrow D^0\pi^+_s\rightarrow K^0_S\eta\pi^+_s $&~~$45831\pm283$ &$+1.00\pm0.51$\\ 
$D^{*+}\rightarrow D^0\pi^+_s\rightarrow K^0_S\eta'\pi^+_s$&~~$26899\pm211$ &$+1.47\pm0.67$\\ 
    \end{tabular}     
  \end{ruledtabular}
\end{table}

In order to obtain $A^{\pi^+_s}_{\epsilon}$ we first extract $A^{\rm
  untagged}_{\rm rec}$ using simultaneous fits analogous to those used for the
  signal modes, but instead of the $M(D^*)-M(D)$ distribution we fit to the
  $M(D)$ distribution using a similar parameterization. The values of $A^{\rm
  untagged}_{\rm rec}$ are evaluated in bins of transverse momentum (${p^{\rm
  lab}_{T{D^0}}}$) and polar angle ($\cos\theta^{\rm lab}_{D^0}$) of untagged
  $D^0\rightarrow K^-\pi^+$ candidates in the laboratory frame. The $p_T$ and
  polar angle variables are only weakly correlated. Each tagged
  $D^{*}\rightarrow D\pi_s\rightarrow K\pi\pi_s$ candidate is then weighted
  with $1-A^{\rm untagged}_{\rm rec}$ for $D^{*+}$ and $1+A^{\rm untagged}_{\rm
  rec}$ for $D^{*-}$. Details of the weighting procedure are described in
  Ref.~\cite{D0hhBelle}. After this the remaining asymmetry in the tagged decay
  sample is $A^{\pi^+_s}_{\epsilon}$, which is obtained from the simultaneous
  fits to the weighted $M(D^*)-M(D)$ distributions with the same
  parameterization used in the signal modes, now for bins of $p^{\rm
  lab}_{T\pi^+_{s}}$ and $\cos\theta^{\rm lab}_{\pi^+_s}$. 

The dominant sources of uncertainty in the $A^{\pi^+_s}_{\epsilon}$
determination are the statistical uncertainties in the untagged and tagged
samples. These are found to be 0.04\% and 0.07\%, respectively. Other sources
of systematic uncertainties are found to be negligible. Thus, we assign a total
systematic uncertainty of 0.08\% to the $A^{\pi^+_s}_{\epsilon}$ determination,
obtained by adding the two contributions in quadrature.

The data samples shown in Fig.~\ref{FIG:MKSP0} are divided into bins of $p^{\rm
  lab}_T$ and $\cos\theta^{\rm lab}$ of the $\pi^+_s$. The
$A^{\pi^+_s}_{\epsilon}$ correction is applied by weighting each $D^{*+}$ event
with $1-A^{\pi^+_s}_{\epsilon}$ and each $D^{*-}$ event with
$1+A^{\pi^+_s}_{\epsilon}$. The weighted mass difference distributions in bins
of the $D^{*+}$ polar angle in the CMS are fitted simultaneously to obtain the
corrected asymmetry. We fit for the linear component in $\cos\theta^{\rm
  CMS}_{D^{*+}}$ to determine $A_{FB}$ while the $A_{CP}$ component is uniform
in $\cos\theta^{\rm CMS}_{D^{*+}}$.
\begin{figure}[htbp]
\mbox{
  \includegraphics[height=0.15\textwidth,width=0.48\textwidth]{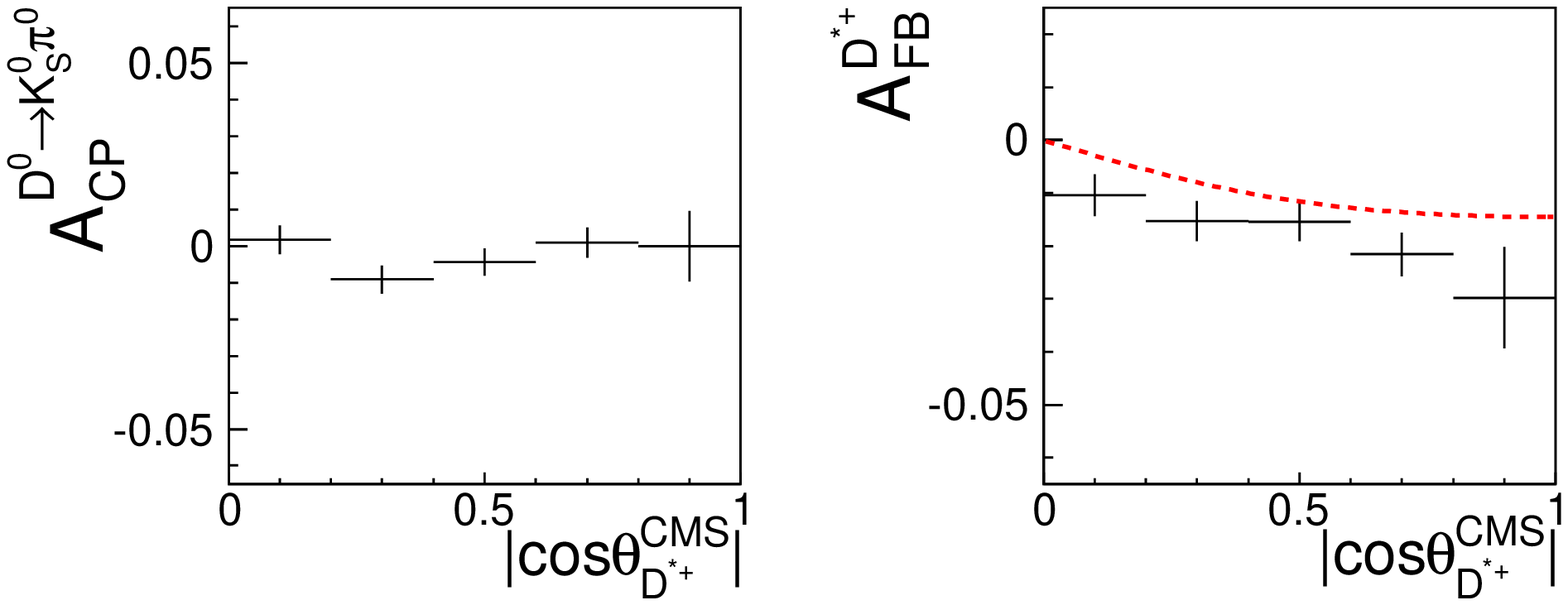}
}
\mbox{
  \includegraphics[height=0.15\textwidth,width=0.48\textwidth]{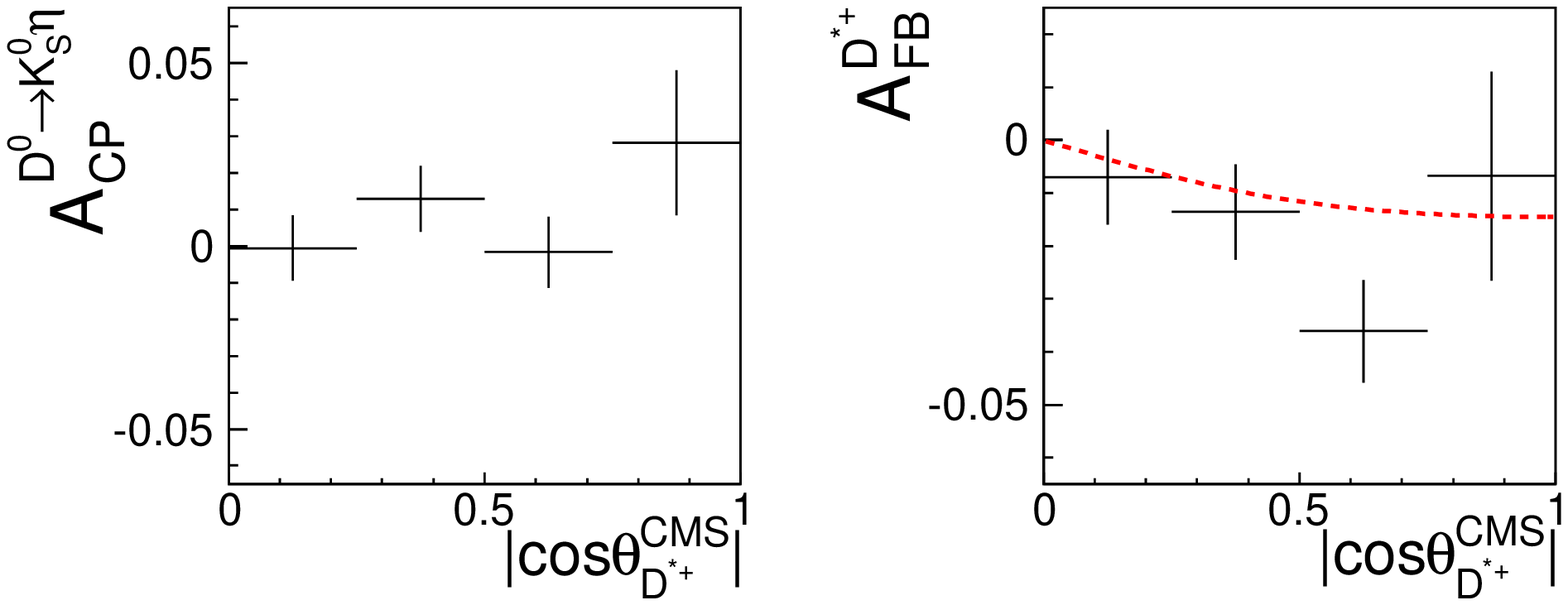}
}
\mbox{
  \includegraphics[height=0.15\textwidth,width=0.48\textwidth]{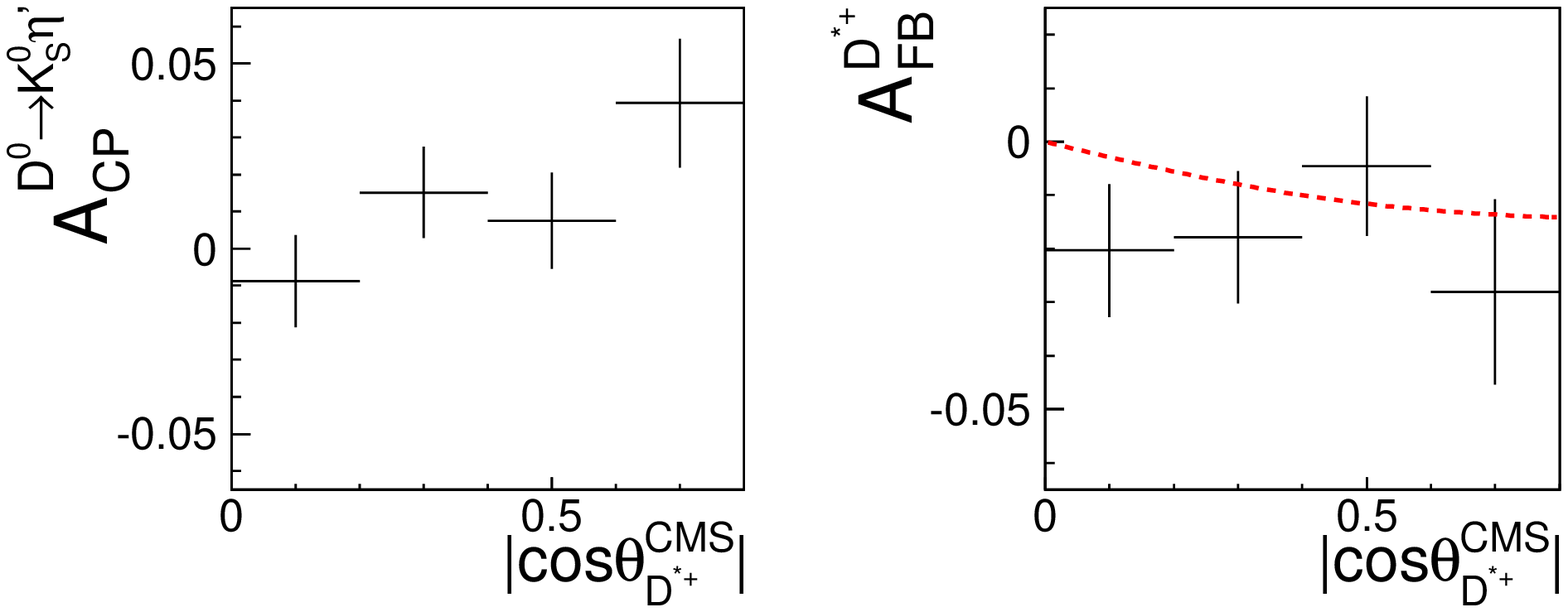}
}
\caption{Measured $A_{CP}$ (left) and $A_{FB}$ (right) values as a function of
  $|\cos\theta^{\rm CMS}_{D^{*+}}|$. Top plots are for $K^0_S\pi^0$, middle
  plots for $K^0_S\eta$, and bottom plots for $K^0_S\eta'$ final states. The
  dashed curves show the leading-order prediction for $A^{c\bar{c}}_{FB}$.}
\label{FIG:ACPD0}
\end{figure}
Figure~\ref{FIG:ACPD0} shows $A^{D^0\rightarrow K^0_S P^0}_{CP}$ and
$A^{D^{*+}}_{FB}$ as a function of $|\cos\theta^{\rm CMS}_{D^{*+}}|$. From a
weighted average over the $|\cos\theta^{\rm CMS}_{D^{*+}}|$ bins, we obtain
$A^{D^0\rightarrow K^0_S\pi^0}_{CP}=(-0.28\pm0.19)\%$, $A^{D^0\rightarrow
  K^0_S\eta}_{CP}=(+0.54\pm0.51)\%$, and $A^{D^0\rightarrow
  K^0_S\eta'}_{CP}=(+0.98\pm0.67)\%$ where the uncertainties are statistical
only. The $\chi^2/{\rm d.o.f}$ with respect to the average over the
$|\cos\theta^{\rm CMS}_{D^{*+}}|$ bins is 5.1/4 $(K^0_S\pi^0)$, 3.0/3
$(K^0_S\eta)$, or 5.3/3 $(K^0_S\eta')$. The observed $A_{FB}$ values decrease
with $\cos\theta^{\rm CMS}_{D^{*+}}$ as expected from the leading-order
prediction~\cite{AFBCC}. The observed deviations from the prediction are
expected due to higher order corrections. Similar $A_{FB}$'s were found in
previous measurements~\cite{D0hhBelle,D0hhBaBar,ACPKSH}. The results are
validated with toy pseudo-experiments and full detector simulation Monte Carlo
events. We found no systematic deviations from the input values.

We consider several sources of systematic uncertainty. The uncertainty due to
the limited size of the tagged and untagged samples was discussed above. To
estimate the systematic uncertainty due to the choice of fitting method and
parameters, we vary the histogram binnings, fitting intervals, and signal and
background parameterizations. We also consider the systematic uncertainties due
to the choice of $\cos\theta^{\rm CMS}_{D^{*+}}$ binning. Finally, we include
possible effects due to the differences in interactions of $K^0$ and
$\bar{K}^0$ mesons with the material of the detector as explained in
Ref.~\cite{ACPKSH}, and assign a systematic uncertainty of 0.06\% due to this
effect. Table~\ref{TABLE:SYSTEM} summarizes the components of the systematic
uncertainties. The larger uncertainties in $A^{D^0\rightarrow
  K^0_S\eta^{(\prime)}}_{CP}$ due to the choice of fitting method are a
consequence of smaller statistics of these samples.
\begin{table}[htbp]
\caption{\label{TABLE:SYSTEM} Summary of systematic uncertainties in $A_{CP}$.}
\begin{ruledtabular}
\begin{tabular}{cccc} 
Source                            &$K^0_S\pi^0$ (\%) &$K^0_S\eta$ (\%) &$K^0_S\eta'$ (\%)\\ \hline
$A^{\pi^+_s}_{\epsilon}$ determination    &~~0.08&0.08&0.08\\ 
Fitting                                   &~~0.02&0.12&0.10\\ 
$\cos\theta^{\rm CMS}_{D^{*+}}$ binning   &$<$0.01&0.01&0.03\\ 
$K^0/\bar{K}^0$-material effects          &~~0.06&0.06&0.06\\ \hline
Total                                     &~~0.10&0.16&0.14\\ 
\end{tabular}     
\end{ruledtabular}
\end{table}

From the total uncertainties shown in Table~\ref{TABLE:SYSTEM}, we obtain
$A_{CP}^{D^0\rightarrow K^0_S\pi^0}=(-0.28\pm0.19\pm0.10)\%$,
$A_{CP}^{D^0\rightarrow K^0_S\eta}=(+0.54\pm0.51\pm0.16)\%$ and
$A_{CP}^{D^0\rightarrow K^0_S\eta'}=(+0.98\pm0.67\pm0.14)\%$ where the first
uncertainties are statistical and the second are
systematic. Table~\ref{TABLE:SUMMARY} summarizes the results, current world
average~\cite{PDG2010}, and $A^{\bar{K}^0}_{CP}$.
\begin{table}[htbp]
\caption{Summary of the $A_{CP}$ measurements. The first uncertainties in the
  second column are statistical and the second are systematic. The third column
  shows the world average of $A_{CP}$ and the fourth
  $A^{\bar{K}^0}_{CP}$. $A^{D^0\rightarrow K^0_S\eta^{(\prime)}}_{CP}$ are the
  first measurements, hence no world average of $A_{CP}$ is given in the third
  column.}
\label{TABLE:SUMMARY}
\begin{ruledtabular}
\begin{tabular}{cccc} 
                                    &Belle (\%)&Ref.~\cite{PDG2010} (\%) &$A^{\bar{K}^0}_{CP}$ (\%) \\ \hline
$A^{D^0\rightarrow K^0_S\pi^0}_{CP}$&$-0.28$$\pm$$0.19$$\pm0.10$&$+0.1$$\pm$$1.3$ &$-0.332$$\pm$$0.006$\\ 
$A^{D^0\rightarrow K^0_S\eta}_{CP}$ &$+0.54$$\pm$$0.51$$\pm0.16$&---&$-0.332$$\pm$$0.006$\\ 
$A^{D^0\rightarrow K^0_S\eta'}_{CP}$&$+0.98$$\pm$$0.67$$\pm0.14$&---&$-0.332$$\pm$$0.006$\\ 
\end{tabular}     
\end{ruledtabular}
\end{table}

Besides the $A_{CP}$ measurements listed in Table~\ref{TABLE:SUMMARY}, we test
the universality of $a^{\rm ind}$ assuming negligible new $CP$ violating
effects in $D^0$ decays to the $K^0_S\pi^0$ final state as discussed in
Ref.~\cite{YAY}. By subtracting $A^{\bar{K}^0}_{CP}$ from $A^{D^0\rightarrow
  K^0_S\pi^0}_{CP}$, we obtain $a^{\rm ind}=(+0.05\pm0.19\pm0.10)\%$, which is
consistent with $-A_{\Gamma}=(-0.01\pm0.30\pm0.15)\%$ obtained in
Ref.~\cite{D0MIXING_Belle}. This is the first experimental test of $a^{\rm
  ind}$ in $D^0$ decays with a sensitivity near 0.3\%. By averaging the two
independent values we obtain $a^{\rm ind}=(+0.03\pm0.18)\%$, where the
uncertainty includes the statistical and systematic errors, and represents the
most precise value of $a^{\rm ind}$ from a single-experiment currently. Using
the average $a^{\rm ind}$, we also update the values of $a^d_{D^0\rightarrow
  K^+K^-}$ and $a^d_{D^0\rightarrow\pi^+\pi^-}$ from Ref.~\cite{D0hhBelle},
which are $(-0.46\pm0.37)\%$ and $(+0.40\pm0.56)\%$~\cite{CORR_AB},
respectively. The errors include all the uncertainties of input measurements.

In summary, we report a search for $CP$ violation in the decays $D^0\rightarrow
K^0_S P^0$ using a data sample with an integrated luminosity of 791 fb$^{-1}$
collected with the Belle detector. We observe no evidence for $CP$
violation. The measurement in the decay $D^0\rightarrow K^0_S\pi^0$ is the most
precise measurement of any $CP$ asymmetry in the charmed particle sector to
date. We also report the first measurements of $CP$ asymmetries in the decays
$D^0\rightarrow K^0_S\eta$ and $D^0\rightarrow K^0_S\eta'$. Our results are
consistent with the SM and can be used to place the most stringent constraints
on NP models arising from the measurements of $CP$ violation in the charm
sector at present. 

We thank the KEKB group for excellent operation of the accelerator, the KEK
cryogenics group for efficient solenoid operations, and the KEK computer group
and the NII for valuable computing and SINET3 network support. We acknowledge
support from MEXT, JSPS and Nagoya's TLPRC (Japan); ARC and DIISR (Australia);
NSFC (China); MSMT (Czechia); DST (India); MEST, NRF, NSDC of KISTI, and WCU
(Korea); MNiSW (Poland); MES and RFAAE (Russia); ARRS (Slovenia); SNSF
(Switzerland); NSC and MOE (Taiwan); and DOE (USA). B.~R.~Ko acknowledges
support by NRF Grant No. 2010-0021279 and E.~Won acknowledges support by NRF
Grant No. 2009-0071072.


\begin{thebibliography}{99}

\bibitem{D0MIXING_Babar}
B. Aubert {\it et al.} (Babar Collab.), Phys. Rev. Lett. {\bf 98}, 211802 (2007).

\bibitem{D0MIXING_Belle}
M. Stari{\v c} {\it et al.} (Belle Collab.), Phys. Rev. Lett. {\bf 98}, 211803 (2007).

\bibitem{D0MIXING_CDF}
T. Aaltonen {\it et al.} (CDF Collab.), Phys. Rev. Lett. {\bf 100}, 121802 (2008).

\bibitem{HFAG} 
D. Asner {\it et al.} (Heavy Flavor Averaging Group), \url{arXiv:1010.1589v1 [hep-ex]} and online update at \url{http://www.slac.stanford.edu/xorg/hfag/}.

\bibitem{DMIX} 
A. F. Falk, Y. Grossman, Z. Ligeti, and A. A. Petrov, Phys. Rev. D {\bf 65}, 054034 (2002);
A. F. Falk, Y. Grossman, Z. Ligeti, Y. Nir, and A. A. Petrov, Phys. Rev. D {\bf 69}, 114021 (2004).

\bibitem{YAY}
Y. Grossman, A. L. Kagan, and Y. Nir, Phys. Rev. D {\bf 75}, 036008 (2007).

\bibitem{SMCP}
F. Buccella, M. Lusignoli, G. Miele, A. Pugliese, and P. Santorelli, Phys. Rev. D {\bf 51}, 3478 (1995).
  
\bibitem{CC}
Throughout this Letter the charge-conjugate decay mode is also implied unless
stated otherwise.

\bibitem{PDG2010}
K. Nakamura {\it et al.} (Particle Data Group), J. Phys. G {\bf 37}, 075021 (2010).

\bibitem{ACPK0S}
We use
$A^{\bar{K}^0}_{CP}~=~\frac{\Gamma(\bar{K}^0\rightarrow\bar{f})-\Gamma(K^0\rightarrow
  f)}{\Gamma(\bar{K}^0\rightarrow\bar{f})+\Gamma(K^0\rightarrow f)}$. Hence,
$A^{\bar{K}^0}_{CP}=-A_{L}$ where $A_{L}$ is the symbol for the asymmetry used
in Ref.~\cite{PDG2010}.

\bibitem{KM}
M. Kobayashi and T. Maskawa, Prog. Theor. Phys., {\bf 49}, 652 (1973).

\bibitem{BIGI}
I. I. Bigi and H. Yamamoto, Phys. Lett. B {\bf 349}, 363 (1995).

\bibitem{D0hhBelle} 
M. Stari{\v c} {\it et al.} (Belle Collab.), Phys. Lett. B {\bf 670}, 190 (2008).

\bibitem{D0hhBaBar} 
B. Aubert {\it et al.} (Babar Collab.), Phys. Rev. Lett. {\bf 100}, 061803 (2008).

\bibitem{BELLE}
A. Abashian {\it et al.} (Belle Collab.), Nucl. Instr. and Meth. A {\bf 479}, 117 (2002).

\bibitem{KEKB}
S. Kurokawa and E. Kikutani, Nucl. Instr. and Meth. A {\bf 499}, 1 (2003), and
other papers included in this volume.

\bibitem{BRKS}
E. Won {\it et al.} (Belle Collab.), Phys. Rev. D {\bf 80}, 111101(R) (2009).

\bibitem{SVD2} 
Z. Natkaniec {\it et al.} (Belle SVD2 Group), Nucl. Instr. and Meth. A {\bf
  560}, 1 (2006); Y. Ushiroda (Belle SVD2 Group), Nucl. Instr. and Meth. A {\bf
  511}, 6 (2003).

\bibitem{ECL}
H. Ikeda {\it et al.}, Nucl. Instr. and Meth. A {\bf 441}, 401 (2000).

\bibitem{AFBCC} 
The leading-order prediction for $A^{c\bar{c}}_{FB}$ at $\sqrt{s}$=10.6 GeV is
about $-0.029\cos\theta^{\rm CMS}/(1+\cos^2\theta^{\rm CMS})$. See for example
O. Nachtmann, Elementary Particle Physics, Springer-Verlag (1989).

\bibitem{ACPKSH} 
B. R. Ko {\it et al.} (Belle Collab.), Phys. Rev. Lett. {\bf 104}, 181602 (2010).

\bibitem{CORR_AB} 
The systematic uncertainty due to the $A^{\pi^+_s}_{\epsilon}$ determination in
the two measurements are only partially correlated due to the use of slightly
different data set and different tracking algorithms. The difference in the
results when treating the systematic uncertainties as completely correlated or
completely uncorrelated is negligible. To be conservative, we quote the result
treating the systematic uncertainties as uncorrelated.

\end{thebibliography}
\end{document}